\documentclass{article}

\usepackage{arxiv}

\usepackage[utf8]{inputenc} 
\usepackage[T1]{fontenc}    
\usepackage{hyperref}       
\usepackage{url}            
\usepackage{booktabs}       
\usepackage{amsfonts}       
\usepackage{nicefrac}       
\usepackage{microtype}      
\usepackage{lipsum}		
\usepackage{graphicx}
\usepackage{natbib}
\usepackage{doi}

\title{COVID-Datathon: Biomarker identification for COVID-19 severity based on BALF scRNA-seq data}

\author{ \href{}{Seyednami Niyakan} \\
	Department of Electrical \& Computer Engineering\\
	Texas A\&M University\\
	College Station, U.S. \\
	\texttt{naminiyakan@tamu.edu} \\
	\And
	\href{}{Xiaoning Qian} \\
	Department of Electrical \& Computer Engineering\\
	Texas A\&M University\\
	College Station, U.S. \\
	\texttt{xqian@ece.tamu.edu} \\
}

\renewcommand{\shorttitle}{COVID-19 Severity Biomarker Identification}

\hypersetup{
pdftitle={A template for the arxiv style},
pdfsubject={q-bio.NC, q-bio.QM},
pdfauthor={David S.~Hippocampus, Elias D.~Striatum},
pdfkeywords={First keyword, Second keyword, More},
}

\begin{document}
\maketitle

\begin{abstract}
	The severe acute respiratory syndrome coronavirus 2 (SARS-CoV-2) emergence began in late 2019 and has since spread rapidly worldwide. The characteristics of respiratory immune response to this emerging virus is not clear. Recently, Single-cell RNA sequencing (scRNA-seq) transcriptome profiling of Bronchoalveolar lavage fluid (BALF) cells has been done to elucidate the potential mechanisms underlying in COVID-19. With the aim of better utilizing this atlas of BALF cells in response to the virus, here we propose a bioinformatics pipeline to identify candidate biomarkers of COVID-19 severity, which may help characterize BALF cells to have better mechanistic understanding of SARS-CoV-2 infection. The proposed pipeline is implemented in R and is available at \url{https://github.com/namini94/scBALF_Hackathon}.
\end{abstract}


\section{Introduction}
Globally, the outbreak of Coronavirus disease 2019 (COVID-19), caused by severe acute respiratory syndrome coronavirus 2 (SARS-CoV-2), has led to more than 233 million infections and more than 4.7 million deaths
according to the statistics of World Health Organization (WHO) as of October 1, 2021~\citep{b2}. The clinical spectrum of COVID-19 is broad: Many COVID-19 patients are asymptomatic or experience only mild symptoms; however, some patients progress to severe life-threatening conditions~\citep{b2}. Accurate classification of COVID-19 severity in patients may aid in delivering proper healthcare
and reducing mortality. Thus it is of great importance to understand the underlying molecular mechanisms of the disease and identify the biomarkers of the illness severity accurately~\citep{b3}. 

Single-cell RNA sequencing (scRNA-seq) is a powerful tool at dissecting the cellular processes and characterizing immune responses \citep{b2,b4}. Many groups have been applying scRNA-seq to COVID-19 studies to better understand the human immune response to the infection~\citep{b2,b5,b6}. A recent study performed scRNA-seq on Bronchoalveolar lavage fluid (BALF) cells of patients with different COVID-19 severity levels to characterize the respiratory immune properties associated with COVID-19 severity \citep{b7}.

To help better understand the inherent biological signals in the recently published single-cell RNA-seq data from BALF cells~\citep{b7} for COVID-19 severity prediction, we develop a bioinformatics pipeline to first identify the COVID-19 severity biomarker genes and then classify BALF cells using these genes. We apply several classification algorithms, including linear classifiers, such as Linear Discriminant Analysis (LDA), and non-linear ones, such as Quadratic and Flexible  Discriminant Analysis (QDA and FDA~\citep{b8}), Random Forests (RF)~\citep{b9} and Support Vector Machines (SVM), to evaluate our pipeline in classifying BALF cells based on COVID-19 severity.

\section{Methods}
\subsection{Data Set}
\begin{figure*}[!t]
    \centering 
        \includegraphics[width=\textwidth,keepaspectratio]{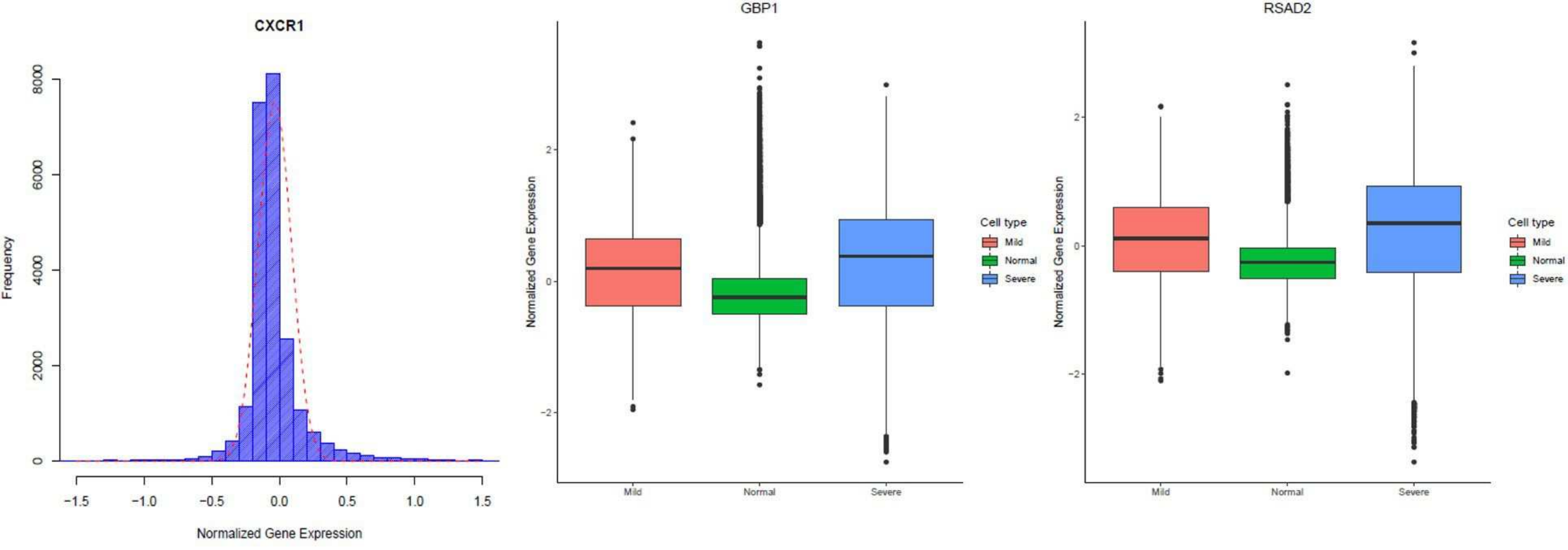}\vspace{-1mm}
        \caption{\textbf{Left:} Histogram of normalized expression values for a random gene "CXCR1" across cells and an approximate Normal distribution shown with red dashed lines. \textbf{Middle \& Right:} Boxplots of normalized expression values for two of identified potential COVID-19 biomarkers, genes "GBP1" and "RSAD2", across different group of cells.  }
        \label{Fig:DE}
\end{figure*}

We have used the scBALF-COVID-19 dataset prepared for the single-cell transcriptomics challenge of the IEEE COVID-19 Data Hackathon. This dataset is derived from public data \citep{b7} and contains a set of Broncho Alveolar Lavage Fluid (BALF) cells from patients categorized clinically as having mild, severe or no COVID-19 infection. In detail, the dataset has scRNA-seq data of 23189 BALF cells from three classes based on COVID-19 severity: mild infection (3292 cells), severe infection (7919 cells), and no infection (11978 cells) across 1999 genes. The dataset has been normalized for technical differences between patients (batch normalization) as well as sequencing depth differences between cells in each patient. This normalization step, has removed the overdispersion inherent in typical scRNA-seq data. As a result, it may not be beneficial to apply typical scRNA-seq tools for biomarker identification and cell clustering on this normalized data set, such as scVI~\citep{b10}, DESeq2~\citep{b11}  and SimCD~\citep{b4}, in which Negative Binomial (NB) distribution models and their extensions were developed to model scRNA-seq data. As shown in the histogram plot of normalized gene expression values for a random gene ``CXCR1'' in the left plot of Fig~\ref{Fig:DE}, normalized gene expression profiles across samples can be approximated with a normal distribution.

\subsection{Biomarker Identification}\label{2B}

At the first step of our proposed pipeline, we detect genes that are differentially expressed (DE) across cells with different levels of COVID-19 severity. In order to do this we perform three different sets of DE analyses, each time as one label vs the rest (e.g. severe cells vs other cells). DE analyses are based on the R package \texttt{Monocle}~\citep{b12} designed specifically for scRNA-seq experiments with the capability of handling normally distributed data. This can be done by passing the argument \texttt{expressionFamily} to be \texttt{uninormal}. In detail, what Monocle does for DE analyses is fitting vector generalized linear models (VGLM) to both full and reduced models and then computing likelihood ratio Chi-squared statistics and corresponding p-values.  After that, for each of three analyses, we rank genes based on their adjusted p-values in an increasing order. Then we report the intersection of top 100 DE genes (also having lower adjusted p-values) in all three DE analyses as COVID-19 severity potential biomarkers. 

\begin{table*}[!ht]
\centering
\caption{{AUC-ROC of classifying cells for one label vs rest using Machine Learning classification algorithms}
\label{Tab:clusZINB}}
\resizebox{1.00\columnwidth}{!}{
{\begin{tabular}{@{}c|c|ccccc@{}}
\toprule {\bf Gene Set} & {\bf Class}  & {\bf LDA} & {\bf QDA} & {\bf FDA} & {\bf SVM (RBF)} & {\bf{RF}} \\\hline \hline
{} & {Sev vs Rest} & {0.7644 $\pm$ 0.0012} &  0.9615 $\pm$ 0.0012 & 0.9778 $\pm$ 0.0007 & 0.9838 $\pm$ 0.0008 & \bf{0.9912 $\pm$ 0.0003} \\
{\bf G1} & {Nor vs Rest} & {0.7363 $\pm$ 0.0008}  & 0.9286 $\pm$ 0.0009  & 0.9387 $\pm$ 0.0008 & 0.9632 $\pm$ 0.0010 & \bf{0.9782 $\pm$ 0.0005} \\
{} & {Mil vs Rest} & {0.6201 $\pm$ 0.0007}  &  0.6882 $\pm$ 0.0013 & 0.8256 $\pm$ 0.0011 & 0.8670 $\pm$ 0.0009 & \bf{0.9160 $\pm$ 0.0006}\\ \hline
{} & {Sev vs Rest} & {0.8777 $\pm$ 0.0011}   &  0.9771 $\pm$ 0.0009 & 0.9992 $\pm$ 0.0005 & 0.9952 $\pm$ 0.0005 & \bf{0.9999 $\pm$ 0.0001}\\
{\bf G2} & {Nor vs Rest} & {0.8350 $\pm$ 0.0009}   &  0.9092 $\pm$ 0.0006 & 0.9778 $\pm$ 0.0008  & 0.9367 $\pm$ 0.0008 & \bf{0.9838 $\pm$ 0.0005} \\
{} & {Mil vs Rest} & {0.6446 $\pm$ 0.0010}  &  0.7311 $\pm$ 0.0008 &  0.9369 $\pm$ 0.0008 & 0.8858 $\pm$ 0.0004 & \bf{0.9627 $\pm$ 0.0005} \\ \hline
{} & {Sev vs Rest} & {0.8888 $\pm$ 0.0007}  &  0.9881 $\pm$ 0.0006 & 0.9998 $\pm$ 0.0002 & 0.9999 $\pm$ 0.0001 & \bf{0.9999 $\pm$ 0.0001}  \\
{\bf G3} & {Nor vs Rest} & {0.8475 $\pm$ 0.0023}&  0.9205 $\pm$ 0.0008 & 0.9965 $\pm$ 0.0012 & 0.9913 $\pm$ 0.0004 & \bf{0.9957 $\pm$ 0.0003} \\
{} & {Mil vs Rest} & 0.6917 $\pm$ 0.0004  & {0.8263 $\pm$ 0.0007} & 0.9917 $\pm$ 0.0002 & 0.9889 $\pm$ 0.0008 & \bf{0.9925 $\pm$ 0.0004}\\ \hline

\end{tabular}}{}
\label{Tabel:ROC}
}
\end{table*}

\subsection{Cell Classification Algorithms}\label{2C}

After identifying the candidate genes as biomarkers to classify BALF cells for COVID-19 infection severity, we run various well-known classifiers, including both linear and non-linear classifiers. We first run LDA as it assumes that predictors are normally distributed and that the different classes have class-specific means and equal variance. Then we try multiple non-linear classifiers, including QDA, FDA, RF and SVM with the radial basis function (RBF) kernel to better capture the non-linearity in the data. For LDA and QDA, we use \texttt{lda} and \texttt{qda} from the R package \texttt{MASS}. For FDA, we use \texttt{fda} with the regression method to be multivariate
adaptive regression splines (MARS), from R package \texttt{mda}. For RF, we use \texttt{randomForest} function from a R package with the same name. We perform SVM with RBF, with the \texttt{svm} from the R package \texttt{e1071}. Also, it worth mentioning that in order to account for the imbalance in class labels we set the option \texttt{class.weights} to be \texttt{inverse} so that weights for samples with different class labels be chosen inversely proportional to the corresponding class size.

\section{Results}

In this section, we present the results of applying our bioinformatics pipeline discussed in Sections \ref{2B} and \ref{2C} to the BALF cell scRNA-seq data. First, we discuss the biomarker identification results and then go over the results of cell classification tools.

\subsection{Biomarker Identification}

After applying the DE analyses detailed in Section \ref{2B}, we end up having 12 genes in the intersection set of top 100 DE genes in three sets of DE analyses performed by Monocle: \emph{RSAD2}, \emph{CXCL10}, \emph{IDO1}, \emph{GCH1}, \emph{CXCL11}, \emph{CRYBA4}, \emph{CCL3}, \emph{LGMN}, \emph{IFIT1}, \emph{CTSB}, \emph{GBP1} and \emph{CCL2}. Half of these genes (\emph{RSAD2}, \emph{CXCL10}, \emph{CXCL11}, \emph{CCL3}, \emph{IFIT1}, \emph{CCL2}) have been previously reported as genetic biomarkers for COVID-19 severity in a recent study that profiled the immune response signatures in the BALF cells of eight COVID-19 cases \citep{b13}. The rest of the genes detected by our pipeline can be new potential biomarkers for immune response to COVID-19 in BALF cells. The middel and right box plots in Fig.~\ref{Fig:DE} show the normalized gene expression values of \emph{GBP1} and \emph{RSAD2}, two identified potential biomarkers by our pipeline. \emph{RSAD2} has been previously studied to be associated with respiratory immune response severity in BALF cells but \emph{GBP1} can be a new potential biomarker.    

\subsection{Cell classification}

After having the ranking lists of genes based on their adjusted p-values in three sets of one class vs rest DE analyses, we make three different sets of candidate gene features for cell classification: 
\begin{enumerate}
    \item \textbf{G1:} Intersection of top 100 DE genes in all three analyses,
    \item \textbf{G2:} Intersection of top 100 DE genes in Normal cells vs rest and Severe cells vs rest DE analyses,
    \item \textbf{G3:} Union of top 100 DE genes in all three analyses.
\end{enumerate}

\begin{figure*}[!t]
    \centering 
        \includegraphics[width=\textwidth,keepaspectratio]{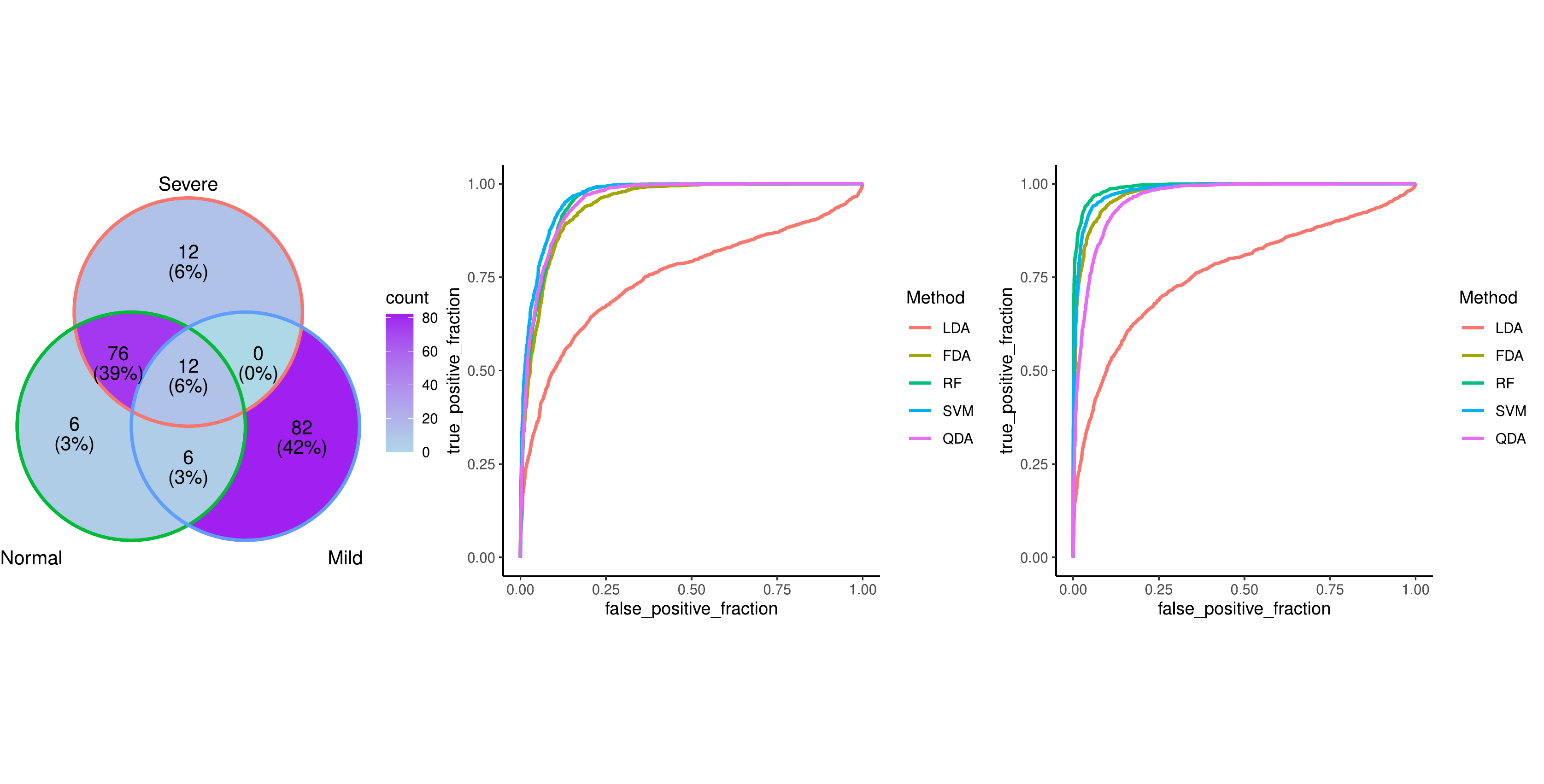}\vspace{-15mm}
        \caption{\textbf{Left:} Venn Diagram of top 100 DE genes from the three sets of DE analyzes: Normal vs rest, Severe vs rest and Mild vs rest. \textbf{Middle \& Right:} ROC curves of different classification algorithms on gene set \textbf{G1} when testing on classifying cells based on Normal vs rest and Severe vs rest.(Corresponding to AUC-ROC values in top two rows of Table.1)  }
        \label{Fig:2}
\end{figure*}

Applying these to our BALF scRNA-seq data, we find 12, 88 and 194 genes in the gene sets \textbf{G1}, \textbf{G2} and \textbf{G3} respectively. Venn diagram of top 100 DE genes from the three sets of DE analyzes is shown in left plot of Figure \ref{Fig:2}. We then apply the previously described classification algorithms in Section \ref{2C} with the gene expression values from each of the gene sets to classify cells. In order to measure the accuracy of each classifier by each of three gene sets, we first randomly split data to $75\%$ and $25\%$ of the total size of the available cell samples to construct training and test sample sets. Then for each gene set and each classifier, we train the classifier three times (each time one label vs the rest) on the training set and calculate area under Receiver Operator Characteristic (AUC-ROC) values by testing the trained model on the test data. In each case of label comparison, classification algorithm and gene set, we run the code three times and then report the average and standard deviations of AUC-ROC values.

Table \ref{Tabel:ROC} shows these AUC-ROC values for each of combinations of running one of five different classifiers on three gene sets. As it was expected, the linear classifier LDA has the worst performance in comparison with non-linear ones. Furthermore, as we expect, almost for all classifiers the AUC-ROC is higher when using the gene set \textbf{G3} comparing to other two gene sets \textbf{G1} and \textbf{G2}. Similar trends when comparing the results using  \textbf{G2} vs \textbf{G1}. This is reasonable as the numbers of genes (biomarkers/features) in the gene sets \textbf{G2} and \textbf{G3} are 7 and 16 times higher than the number of genes in the gene set \textbf{G1}, respectively. Actually, the performance of RF, SVM with the RBF kernel, and FDA on the gene set \textbf{G1}, which only has 12 genes, show the practicality of our proposed biomarker identification approach to detect COVID-19 severity in BALF cells. Table \ref{Tabel:ROC}, also presents the superior performance of the random forest classification algorithm on dissecting BALF cells based on their COVID-19 severity. Also, it worth mentioning that one can almost accurately classify BALF cells, by running the RF algorithm on the proposed set of genes from the biomarker identification step in our bioinformatics pipeline. 

\section{Conclusions}
Due to the outbreak of COVID-19 infection across the world and daily increasing number of mortality in the patients having the infection, it is of great importance to better understand the underlying biological mechanisms in human immune response to this infection. As a result of this, here, we presented a bioinformatics pipeline, capable of first, identifying potential COVID-19 severity biomarkers in BALF cells and secondly, accurately classifying cells based on the constructed set of genes. The proposed pipeline is implemented in R and all the codes are available in the Github page: \url{https://github.com/namini94/scBALF_Hackathon}.

\bibliographystyle{unsrtnat}
\bibliography{references}  






\end{document}